\documentclass[11pt]{article}
\usepackage[utf8]{inputenc}
\usepackage[english]{babel}
\usepackage{import}
\usepackage{tikz}
\usepackage{amsthm,todonotes}
\usepackage{amsmath}
\usepackage{amssymb}
\usepackage{float}
\usepackage{geometry}
\usepackage[hidelinks]{hyperref}
\usepackage{xcolor}
\usepackage{lipsum}
\usepackage{cuted}
\usepackage{framed}
\usepackage{authblk}

\usepackage{thmtools}
\usepackage{thm-restate}
\declaretheorem[name=Theorem,numberwithin=section]{thm}

\newtheorem{redrule}{Reduction PVC}
\newtheorem{proposition}{Proposition}
\newtheorem{corollary}{Corollary}[thm]
\newtheorem{lemma}[thm]{Lemma}
\newtheorem{definition}{Definition}[section]

\newcommand{\defparproblem}[4]{
\vspace{1mm}
\noindent\fbox{
  \begin{minipage}{0.96\textwidth}
  \begin{tabular*}{\textwidth}{@{\extracolsep{\fill}}lr} #1  & {\bf{Parameter:}} #3 \\ \end{tabular*}
  {\bf{Input:}} #2  \\
  {\bf{Objective:}} #4
  \end{minipage}
}\vspace{1mm}}

\newcommand{\WOH}{\textrm{\textup{W[1]-hard}}}
\newcommand{\WTH}{\textrm{\textup{W[2]-hard}}}

\title{Partial Vertex Cover on Graphs of Bounded Degeneracy}
\author[1]{Fahad Panolan}
\author[2]{Hannane Yaghoubizade}
\affil[1]{Department of Computer Science and Engineering, IIT Hyderabad, India (\texttt{fahad@cse.iith.ac.in})}
\affil[2]{Department of Mathematical Sciences, Sharif University of Technology, Iran (\texttt{h.yaghoubizade99@sharif.ir})}
\date{}

\begin{document}

\maketitle

\begin{abstract}
In the {\sc Partial Vertex Cover (PVC)} problem, we are given an $n$-vertex graph $G$ and a positive integer $k$, and the objective is to find a vertex subset $S$ of size $k$ maximizing the number of edges with at least one end-point in $S$. This problem is \WOH\ on general graphs, but admits a parameterized subexponential time algorithm with running time $2^{O(\sqrt{k})}n^{O(1)}$ on planar and apex-minor free graphs [Fomin et al. (FSTTCS 2009, IPL 2011)], and a $k^{O(k)}n^{O(1)}$ time algorithm on bounded degeneracy graphs [Amini et al. (FSTTCS 2009, JCSS 2011)]. Graphs of bounded degeneracy contain many sparse graph classes like planar graphs, $H$-minor free graphs, and bounded tree-width graphs (see Figure~\ref{fig:sparse}). In this work, we prove the following results:

\begin{itemize}
    \item There is an algorithm for {\sc PVC} with running time $2^{O(k)}n^{O(1)}$ on graphs of bounded degeneracy which is an improvement on the previous $k^{O(k)}n^{O(1)}$ time algorithm by Amini et al.~\cite{amini}
    \item {\sc PVC} admits a polynomial compression on graphs of bounded degeneracy, resolving an open problem posed by Amini et al.~\cite{amini}
\end{itemize}
\end{abstract}

\section{Introduction}
 In a covering problem, we are given a family ${\cal F}$ of subsets of a universe $U$, and the objective is to find a minimum size subfamily of ${\cal F}$ covering all the elements in $U$. Well known examples of covering problems are {\sc Set Cover}, {\sc Vertex Cover}, {\sc Dominating Set}, {\sc Facility Location}, {\sc $k$-Median}, {\sc $k$-Center}, etc. Covering problems are fundamental in combinatorial optimization and they are very well studied in all areas of algorithms and complexity.

Another variant of covering problems is partial covering problems. In a partial covering problem, the input is a family ${\cal F}$ of subsets of a universe $U$ and a positive integer $k$. The objective is to find a $k$ size subset of ${\cal F}$ that covers the maximum number of elements in $U$. 
Two prominent examples of partial covering problems on graphs are {\sc Partial Vertex Cover (PVC)} and {\sc Partial Dominating Set (PDS)}, which has got considerable attention in the field of parameterized complexity\footnote{For basic definitions related to parameterized algorithms and complexity we refer to Section~\ref{sec:PC}}.    

\defparproblem{{\sc Partial Vertex Cover (PVC)}}{An undirected graph $G$ and a positive integer $k$}{$k$}{Find a vertex subset $S$ of size $k$ such that the number of edges with at least one end-point in $S$ is maximized}

\defparproblem{{\sc Partial Dominating Set (PDS)}}{An undirected graph $G$ and a positive integer $k$}{$k$}{Find a vertex subset $S$ of size $k$ such that the size of the closed neighborhood of $S$ is maximized}
\\

Even though there are many works on {\sc PVC} and {\sc PDS} in the realm of parameterized complexity, there are still some open questions about these problems. It is previously known that {\sc PVC} is  \WOH~\cite{w1hard} and {\sc PDS}, as a more general problem of {\sc Dominating Set}, is \WTH. Amini et al.~\cite{amini} proved that {\sc PVC} can be solved in time $k^{O(k)}n^{O(1)}$ in bipartite graphs, triangle free graphs, planar graphs, $H$-minor free graphs (for a fixed $H$), and bounded degeneracy graphs. On planar graphs, they gave faster algorithms with running time $2^{O(k)}n^{O(1)}$ for {\sc PVC} and {\sc PDS}.
Later, Fomin et al.~\cite{subexp} gave parameterized subexponential time algorithms with running time $2^{O(\sqrt{k})}n^{O(1)}$ for {\sc PVC} and {\sc PDS} on planar graphs and apex-minor free graphs. 
Also, unlike {\sc Dominating Set}, which is known to be FPT~\cite{domsetFPT} on bounded degeneracy graphs, {\sc PDS} is  \WOH~\cite{pdsw1} in this class.

\begin{figure}\label{fig:sparse}
    \centering
    \begin{tikzpicture}
  % Dialectics
  \node[draw] (boundedtw) at (0,0) {Bounded Tree-Width};
  \node[draw] (exMinor) at (0,-1) {H-Minor Free};
  \node[draw] (exTopMinor) at (0,-2) {H-Topological Minor Free};
  \node[draw] (boundedexp) at (0,-3) {Bounded Expansion};
  \node[draw] (boundeddegene) at (0,-4) {Bounded Degeneracy};
  \node[draw] (planar) at (-3.5,0) {Planar};
  \node[draw] (boundedgenus) at (-3.5,-1) {Bounded Genus};
  
  \draw[->] (boundedtw) to (exMinor);
  \draw[->] (exMinor) to (exTopMinor);
  \draw[->] (exTopMinor) to (boundedexp);
  \draw[->] (boundedexp) to (boundeddegene);
  \draw[->] (planar) to (boundedgenus);
  \draw[->] (boundedgenus) to (exMinor);
\end{tikzpicture}
    \caption{Inclusion relation between various sparse graph classes.}
\end{figure}
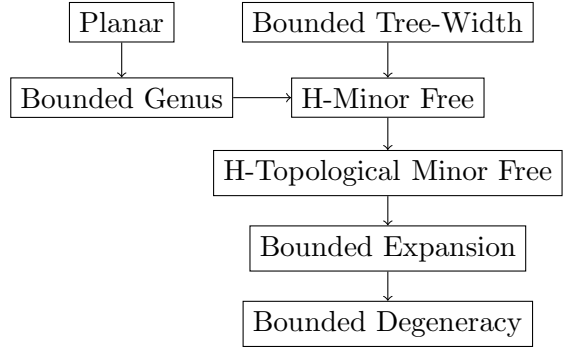

%Recall that Amini et al.~\cite{amini} gave $2^{O(k)}n^{O(1)}$ time algorithms for {\sc PVC} and {\sc PDS} on planar graphs. In~\cite{amini}, Amini et al. wrote: 

%\begin{quote}
     %Finally, let us remark that while {\sc Dominating Set} is FPT on $d$-degenerated graphs~\cite{domsetFPT}, there are strong arguments that our results cannot be extended to this class of sparse graphs. This is because Golovach and Villanger~\cite{pdsw1} showed that {\sc Partial Dominating Set} is \WOH\ on $d$-degenerated graphs.
%\end{quote}

In this work, we give a parameterized single exponential time algorithm for {\sc PVC} on $d$-degenerate graphs. Our algorithm also works for the more general weighted version of the problem.  
\begin{restatable}{thm}{thmalg} \label{thrm:1}
Given $G=(V, E)$, a $d$-degenerate graph with edge weights $w: E\to\mathbb{R}^+$, and an integer $k > 0$, there is an algorithm that runs in $2^{kd + k}(kd)^{O(\log(kd))} n^{O(1)}$ time and finds a subset $S \subseteq V$ of size $k$, with maximum possible $E_G(S)$, i.e., the total weight of edges with at least one end-point in $S$.
\end{restatable}

In~\cite{amini}, Amini et al. asked whether {\sc PVC} and {\sc PDS} admit polynomial kernels on planar graphs. We prove that {\sc PVC} admits a polynomial compression on $d$-degenerate graphs, a more general class of sparse graphs. To get a better size bound for planar graphs, we prove the following general theorem. 
\begin{restatable}{thm}{thmker} \label{thrm:2}
Given a $d$-degenerate graph $G=(V, E)$ that does not contain any $K_{p, p}$ as a subgraph, and an integer $k > 0$, there is a polynomial-time algorithm that outputs a subgraph $H=(V'\subseteq V,E' \subseteq E)$ of $G$ with $O(pd^2(2dk)^{p})$ vertices and a weight function $\rho \colon V' \rightarrow \{0,\ldots,2^{dk}\}$ on the vertex set $V'$ with the following properties. 
\begin{itemize}
\item For any vertex subset $S'\subseteq V' \subseteq V$, $E_G(S')$ is equal to $E_H(S') + \sum_{v\in S'}\rho(v)$.
\item Let $S$ be a partial vertex cover of size $k$ in $G$ covering at least $t$ edges. Then there is a vertex set $S'\subseteq V'$ of size $k$ such that $E_H(S') + \sum_{v\in S'}\rho(v)$ is at least $t$.
%\item If there is a vertex set $S'\subseteq V'$ of size $k$ such that the number of edges incident on $S'$ in $H$ plus $\sum_{v\in S'}\rho(v)$ is at least $t$, then there is a partial vertex cover of size $k$ in $G$ that cover at least $t$ edges.
\end{itemize}
\end{restatable}

%Here, $K_{p,p}$ denote the complete bipartite graph with $p$ vertices on either sides. 
Since planar graphs are $5$-degenerate and do not contain $K_{3, 3}$ as a subgraph, we get the following corollary from Theorem~\ref{thrm:2}.

\begin{corollary}\label{col:1}
There is a polynomial compression for {\sc PVC} on planar graphs. Here, the compressed instance is a graph $H$ with $O(k^3)$ vertices and a weight function on the vertex set of $H$ where the weight of each vertex can be encoded using $O(k)$ bits. 
\end{corollary}

Because no $d$-degenerate graph contains $K_{d+1, d+1}$ as a subgraph, we also have the following corollary.

\begin{corollary}
There is a polynomial compression for {\sc PVC} on $d$-degenerate graphs. Here, the compressed instance is a graph $H$ with $O(k^{d+1})$ vertices and a weight function on the vertex set of $H$, where the weight of each vertex can be encoded using at most $kd$ bits. 
\end{corollary}

\paragraph*{Our methods.}
First, we explain the overview of our FPT algorithm mentioned in Theorem~\ref{thrm:1}, which is based on the following randomized process. Notice that for a $d$-degenerate graph, there is a sequence of vertices such that for any vertex $v$, the number of $v$'s neighbors at the right of it in the sequence is at most $d$. Let $S$ be a solution for {\sc PVC} and let $S'$ be the set of vertices that are not in $S$, but they are a``right neighbor" of a vertex in $S$. Clearly, $|S\cup S'|\leq k+kd$. If we color each vertex red or blue uniformly at random, with probability at least $\frac{1}{2^{k+kd}}$, all the vertices in $S$ would get red, and all the vertices in $S'$ would get blue. Now we assign a value $val(v)$ to any vertex $v$, which is $|N_G(v)|$ minus the number of red ``right neighbors" of $v$. This assignment of values ensures that each edge incident on a red vertex contributes to the value of exactly one red vertex. Observing that for every vertex in $S$ all of its red ``right neighbors" are also in $S$, the solution will be the $k$ most valuable red vertices, and the number of edges covered by them will be the sum of their values. This algorithm can be derandomized using universal sets. In Section~\ref{sec:fpt}, we present the  deterministic version of the algorithm. 

Next, we give a high-level idea of our polynomial compression algorithm. We prove that a ``large" $d$-degenerate graph without any $K_{p,p}$ as a subgraph, has an independent set $I$ of size $k+1$ and a vertex subset $\mathcal{C}$ such that for any distinct $x,y\in I$, $N_G(x)\cap N_G(y) = \mathcal{C}$. Then, we prove that there is a solution that does not contain the least degree vertex of $I$. This leads to a simple reduction rule as long as the number of vertices is not polynomially bounded in $k$. This algorithm is explained in Section~\ref{sec:comp}.

\paragraph*{Other related works.}
In~\cite{ExactAF} some generalization of vertex cover (e.g. {\sc PVC}) parameterizing by tree-width is studied. Also, {\sc PVC} parameterized by the number of covered edges is studied in~\cite{byedge}. There are also extensive works on the approximability of {\sc PVC} on general graphs.~\cite{marx,lossy,Manurangs} For example, Manurangsi in~\cite{Manurangs} presents a simple FPT approximation scheme that runs in $(1/\epsilon)^{O(k)}n^{O(1)}$ as well as an approximation kernelization scheme of $O(k/\epsilon)$ vertices for weighted {\sc PVC}.

%\todo[inline]{Please check the fomin et al paper and add works related to PVC and PDS which are not already mentioned above. Also add FPT approximation and approximate kernelization results here}

%%%%%%%%%%%%%%%%%%%%%%%%%%%%%%%%%%%%%%%%%%%%%%%%%%%%

\section{Preliminaries}
For a graph $G=(V, E)$, we denote the number of vertices and edges by $n$ and $m$, respectively. For a vertex $v\in V$ we denote the set of neighbors of $v$ by $N_G(v)$ and the degree of $v$ by $|N_G(v)|$. For $A \subseteq V$, we use $E_G(A)$ to denote the total number (weight) of edges with at least one end-point in $A$. We denote a complete bipartite graph with partitions of size $p$ and $q$ by $K_{p, q}$.
We use $[n]$ to denote the set $\{1, 2, \dots, n\}$.

\begin{definition}[\textbf{$d$-degenerate graph}]\label{def:1}
An undirected graph $G$ is said to be $d$-degenerate if every subgraph of $G$ contains a vertex of degree at most $d$. The degeneracy of a graph is the smallest value of $d$ for which it is $d$-degenerate.
\end{definition}

We use the following proposition to derive Corollary \ref{col:1} from Theorem \ref{thrm:2}.
\begin{proposition}\label{prop:1}
Planar graphs are $5$-degenerate.
\end{proposition}
\begin{proof}
By Euler's formula, we know $m \leq 3n - 6$ for all $n \geq 3$. Therefore, $\sum_{v \in V} |N_G(v)| \leq 6n-12$, and there is a vertex of a degree at most $5$ in any planar graph. Since every subgraph of a planar graph is also planar, planar graphs are $5$-degenerate.
\end{proof}

For a graph $G=(V, E)$, let $\lambda$ be an ordering of vertices of $G$; i.e. $\lambda: [n] \to V$ is a bijective function. We say $\lambda$ is \textit{d-posterior}, if $\lambda(i)$ has at most $d$ neighbors among $\lambda(i+1), \lambda(i+2), \dots, \lambda(n)$. 
Also, for $v = \lambda(i)$, we call $N_G(v) \cap \{\lambda(i+1), \lambda(i+2), \dots, \lambda(n)\}$ \textit{posterior neighbors} of $v$ and we denote them by $PN_{\lambda}(v)$. Note that since $\lambda$ is a $d$-posterior ordering, we have $PN_{\lambda}(v) \leq d$ for all $v \in V$.
Next, we will state some useful propositions about $d$-degenerate graphs.

\begin{proposition}\label{prop:2}
There exists a $d$-posterior ordering for vertices of any $d$-degenerate graph $G$.
\end{proposition}
\begin{proof}
Let $G_1=G$, and for $2\leq i \leq n$ construct $G_i$ by removing the minimum degree vertex from $G_{i-1}$.
Set $\lambda(i)$ to be a minimum degree vertex in $G_i$.
\end{proof}
\begin{proposition}\label{prop:3}
For a $d$-degenerate graph $G=(V, E)$, we have $m \leq nd$.
\end{proposition}
\begin{proof}
Consider a $d$-posterior ordering $\lambda$ and note that $m = \sum_{v \in V} PN_{\lambda}(v) \leq nd$ because $PN_{\lambda}(v) \leq d$ for any $v \in V$.
\end{proof}
\begin{proposition}\label{prop:4}
Let $G=(V, E)$ be a $d$-degenerate graph. Then, there is a $(d+1)$-coloring for $V$ such that for any $(u, v)\in E$, $u$ and $v$ get different colors; i.e., $f: V \to [d+1]$ such that $f(u) \ne f(v)$ for all $(u, v)\in E$. Furthermore, one can construct this coloring in time $n^{O(1)}$.
\end{proposition}
\begin{proof}
Let $\lambda$ be a $d$-posterior ordering of $V$ and for each $i$ from $n$ to $1$, choose a color for $\lambda(i)$ which does not occur in $PN_{\lambda}(\lambda(i))$.
\end{proof}

\subsection{Parameterized Complexity}
\label{sec:PC}
We state the following definitions slightly modified from the Kernelization book~\cite{kernelization}.
\begin{definition} [\textbf{FPT optimization problem}]
A parameterized optimization problem $\Pi$ is fixed parameter tractable (FPT) if there is an algorithm (called FPT algorithm) that solves $\Pi$, such that the running time of the algorithm on instances of size $n$ with parameter $k$ is upper bounded by $f(k) . n^{O(1)}$ for a computable function $f$.
\end{definition}

\begin{definition} [\textbf{Polynomial-time preprocessing algorithm}]
A polynomial-time preprocessing algorithm $\mathcal{A}$ for a parameterized optimization problem $\Pi$ is a pair of polynomial-time algorithms. The first one is called the \textbf{reduction algorithm}, and given an instance $(I, k)$ of $\Pi$, the reduction algorithm outputs
an instance $(I', k') = \mathcal{R}_\mathcal{A}(I, k)$ of a problem $\Pi'$.
The second algorithm is called the \textbf{solution lifting} algorithm. This algorithm takes an instance $(I, k)$ of $\Pi$, the output
instance $(I', k')$ of the reduction algorithm, and a solution $s'$ to the instance $(I', k')$. The solution lifting algorithm works in time polynomial in $|I|,k,|I'|, k'$ and $|s'|$, and outputs a solution $s$ to $(I, k)$ such that if $s'$ is an optimal solution to $(I', k')$ then $s$ is an optimal solution to $(I, k)$.
\end{definition}
\begin{definition} [\textbf{Compression, Kernelization}]
A polynomial time preprocessing algorithm $\mathcal{A}$ is called a compression, if $\text{size}_\mathcal{A}$ is upper bounded by a computable function $g: \mathbb{N} \to \mathbb{N}$ where $\text{size}_\mathcal{A}$ is defined as follows:
\begin{align*}
    \text{size}_\mathcal{A}(k) = \sup \{|I'| + k': (I', k') = \mathcal{R}_\mathcal{A}(I, k)\ \text{for any instance $(I, k)$ of the problem}\}
\end{align*}
If the upper bound $g(.)$ is a polynomial function of $k$, we say $\mathcal{A}$ is a polynomial compression. 
A compression (polynomial compression) is called a kernelization (polynomial kernelization) if the input and output of the reduction algorithm are instances of the same problem, i.e., $\Pi = \Pi'$.
\end{definition}
%\todo[inline]{define FPT, kenerlizationa and compresssion}

%%%%%%%%%%%%%%%%%%%%%%%%%%%%%%%%%%%%%%%%%%%%%

\section{FPT Algorithm for Weighted Partial Vertex Cover}
\label{sec:fpt}
In this section, we show that PVC can be solved in parameterized single exponential time on  $d$-degenerate weighted graphs. That is, we prove  Theorem \ref{thrm:1}. 

We will use a \textit{universal set} in our algorithm defined as follows. (see also section 5.6.1 of~\cite{parameterized})

\begin{definition}[\textbf{$(n, l)$-universal set}]\label{def:2}
An $(n, l)$-universal set is a family $\mathcal{U}$ of subsets of $[n]$ such that for any $A \subseteq [n]$ of size $l$, the family $\{U \cap A : U \in \mathcal{U}\}$ contains all $2^l$ subsets of $A$.
\end{definition}

\begin{lemma}[Naor et al. \cite{naor}]\label{lem:1}
For any $n,l \geq 1$, one can construct an $(n,l)$-universal set of size $2^ll^{O(\log l)} \log n$ in time $2^ll^{O(\log l)} n\log n$.
\end{lemma}

We now describe our FPT algorithm for solving PVC in the given $d$-degenerate weighted graph $G=(V, E)$. To give a better intuition, we first state the algorithm informally. 
Consider a $d$-posterior ordering for the vertices. Suppose we have an oracle that colors the vertices with blue and red, such that all vertices in the solution get red, all vertices that are not in the solution but are a posterior neighbor of a vertex in the solution get blue, and other vertices get either red or blue. Observe that the solution is a subset of red vertices such that for any vertex in the solution, its red posterior neighbors are also in the solution. Then we will assign a value to each vertex, such that the solution will be the set of $k$ most valuable red vertices. In the algorithm, we use a universal set instead of the oracle. The following is the exact description of the algorithm.

Let $\lambda$ be a $d$-posterior ordering of $V$ and $l = \min(n, k + kd)$. First, we construct an $(n, l)$-universal set $\mathcal{U}$ of subsets of $V$, and for each $U \in \mathcal{U}$ with size $\geq k$ and $v \in V$, we define the \textit{value} of $v$ with respect to $U$ as:
\begin{align*}
    val_U(v) = \sum_{u \in N_G(v)\setminus\left(PN_{\lambda}(v) \cap U\right)} w(u, v)
\end{align*}
And we define $sol(U) \subseteq U$ as the set of $k$ most valuable vertices in $U$, and we set the value of $U$ to be $val(U) = \sum_{v\in sol(U)} val_U(v)$.
Finally, we return $sol(U)$ for the most valuable $U$.

To prove Theorem \ref{thrm:1}, first we show the following lemmas.
\begin{lemma}\label{lem:2}
For any $U\in \mathcal{U}$ and $A \subseteq U$, we have $\sum_{v\in A} val_U(v) \leq E_G(A)$.
\end{lemma}
\begin{proof}
Recall that $E_G(A)$ is the total weight of edges with at least one end-point in $A$.

Any edge $e = (u, v)$ with exactly one end-point, say $v$, in $A$ is counted at most once in $val_U(v)$ and since $u \notin A$, it is also counted at most once in $\sum_{v\in A} val_U(v)$.

For an edge $e'=(u', v')$ with both end-points in $A$, without loss of generality, suppose $u'$ is later than $v'$ in the ordering $\lambda$, i.e., $\lambda^{-1}(u') > \lambda^{-1}(v')$. Therefore, $u' \in PN_{\lambda}(v')$ and since $A \subseteq U$, $u' \in PN_{\lambda}(v') \cap U$ and $e$ is not counted in $val_U(v')$. On the other hand, $v' \notin PN_{\lambda}(u')$, and $e$ is counted in $val_U(u')$. Therefore, $e$ is counted exactly once in $\sum_{v\in A} val_U(v)$.

Since the weights of edges are positive and all edges counted exactly once in $E_G(A)$ are counted at most once in $\sum_{v\in A} val_U(v)$, we have $\sum_{v\in A} val_U(v) \leq E_G(A)$.
\end{proof}

Now, let $S$ be a hypothetical solution, and define $\tilde{S} = S\cup \left(\bigcup_{v\in S} PN_{\lambda}(v)\right)$. Note that:
\begin{align} \label{e1:1}
    |\tilde{S}| \leq & \left|S\right| +\left|\bigcup_{v\in S} PN_{\lambda}(v)\right| \leq k + k.d \tag{since $|S| = k$ and $PN_{\lambda}(v) \leq d$}
\end{align}
Therefore we have $|\tilde{S}| \leq l$. Consider a subset $T \subseteq V$ with size $l$ such that $\tilde{S} \subseteq T$. According to Definition \ref{def:2}, there is a set $\tilde{U} \in \mathcal{U}$ such that $S = \tilde{U} \cap T$. Note that since $|S|=k$, size of $\tilde{U}$ is $\geq k$, and $val_{\tilde{U}}$ and $sol(\tilde{U})$ are defined.
\begin{lemma}\label{lem:3}
$E_G(S) = \sum_{v\in S} val_{\tilde{U}}(v)$.
\end{lemma}
\begin{proof}
It is enough to show that each edge with at least one end-point in $S$ is counted exactly once in $\sum_{v\in S} val_{\tilde{U}}(v)$.

Consider any $e = (u, v)$ with exactly one end-point, say $v$, in $S$. Note that $u \notin S$ and
$$\left(PN_{\lambda}(v) \cap \tilde{U}\right) \subseteq (\tilde{S}\cap \tilde{U})\subseteq (T\cap \tilde{U}) = S$$
Therefore, $u \notin \left(PN_{\lambda}(v) \cap \tilde{U}\right)$ and $e$ is counted in $val_{\tilde{U}}(v)$. Since $u \notin S$, $e$ is counted in $\sum_{v\in S} val_{\tilde{U}}(v)$ exactly once.
For edges with two end-points in $S$, the proof is the same as the proof of Lemma \ref{lem:2}.
\end{proof}

We finally prove Theorem \ref{thrm:1}. For convenience, we restate the theorem here.  

\thmalg*

\begin{proof}
By Lemma \ref{lem:2} and optimality of $S$, $val(U) = \sum_{v\in sol(U)} val_U(v) \leq E_G(sol(U)) \leq E_G(S)$ for all $U \in \mathcal{U}$ with size $\geq k$. Also, note:
\begin{align}
    val(\tilde{U}) &= \sum_{v\in sol(\tilde{U})} val_{\tilde{U}}(v) \tag{definition of $val(\tilde{U})$} \\ 
    &\geq \sum_{v\in S} val_{\tilde{U}}(v) \tag{definition of $sol(\tilde{U})$ and since $S \subseteq \tilde{U}$}\\
    &= E_G(S) \tag{Lemma \ref{lem:3}}
\end{align}
Therefore, for the most valuable $U$, $val(U) = E_G(S)$. Since $val(U) \leq E_G(sol(U)) \leq E_G(S)$, $sol(U)$ is also a solution and $E_G(sol(U)) = val(U)$. This implies the algorithm's correctness and shows that the weight of the edges covered by the solution is equal to $val(U)$.

Finally, the running time of constructing the family $\mathcal{U}$ is $2^{kd + k}(kd + k)^{O(\log (kd + k))} n^{O(1)}$ and more than that, we only have a polynomial process for each $U \in \mathcal{U}$. Since, the size of $\mathcal{U}$ is $2^{kd + k}(kd + k)^{O(\log (kd + k))}\log n$, the total running time is $2^{kd + k}(kd + k)^{O(\log (kd + k))} n^{O(1)}$.
\end{proof}

%%%%%%%%%%%%%%%%%%%%%%%%%%%%%%%%%%%%%%%%%%%%%%%%%%%%%%%%%%%

\section{Polynomial Compression for Partial Vertex Cover}
\label{sec:comp}
In this section, we present a polynomial compression for {\sc PVC} in families of graphs with bounded degeneracy. That is, we prove Theorem~\ref{thrm:2}.

%\import{}{Thrm2.tex}

For convenience we will allow self-loops for this part, but not parallel edges. For a vertex $v$ with self-loops, we will not count $v$ in $N_G(v)$ so $v \notin N_G(v)$, and we will use $L_G(v)$ to denote the number of self-loops of $v$. 
Therefore, the given graph $G=(V, E)$ is undirected, unweighted and any $v \in V$ might have several self-loops. 
Also, $G$ does not contain $K_{p, p}$ and without considering self-loops, it is $d$-degenerate.

\begin{figure}[h]
    \centering 
    \import{}{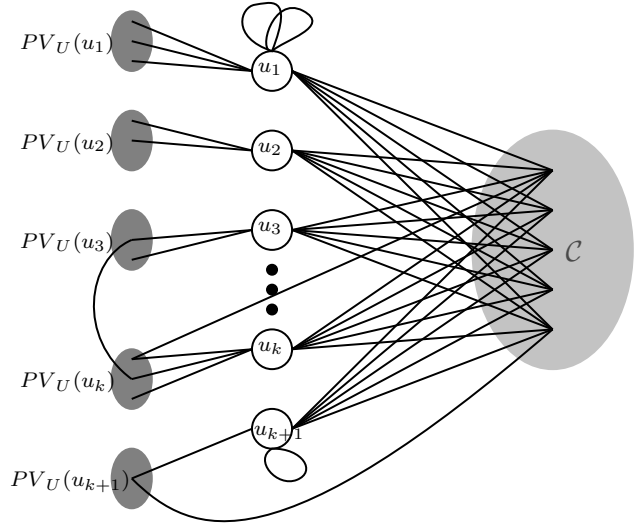}
    \caption{A \textit{nice} subset $U = \{u_1, u_2, \dots, u_{k+1}\}$}
    \label{fig:1}
\end{figure}

We say a subset $U \subseteq V$ of size $k+1$ is \textit{nice} if $U$ is an independent set and there is a subset $\mathcal{C} \subseteq V$ such that for any $u, u' \in U$, $N_G(u) \cap N_G(u') = \mathcal{C}$. For each $u \in U$, we call $N_G(u)\setminus \mathcal{C}$ \textit{private neighbors} of $u$ with respect to $U$, and we denote it by $PV_U(u)$. Figure \ref{fig:1} shows a \textit{nice} subset.

\begin{lemma}\label{lem:4}
Let $G = (V, E)$ be an undirected graph with possible self-loops. For integers $h, p \geq 1$, suppose $I \subseteq V$ is an independent set of size $t > p.(hk)^{p}$, such that $|N_G(v)| \leq h$ for all $v \in I$. Then either there is a \textit{nice} $U \subseteq I$ or $G$ contains a $K_{p, p}$. Furthermore, having $G$ and $I$, we can find a $\textit{nice}$ subset or a $K_{p, p}$ in polynomial time.
\end{lemma}

\begin{proof}
First, we show by induction that for each $0 \leq i \leq p$, either (a) there is a $\textit{nice}$ subset $U \subseteq I$, or (b) there is a $U_i \subseteq I$ of size $t_i > p.(hk)^{p-i}$ such that $Q_i = \bigcap_{u \in U_i} N_G(u)$ has size $\geq i$.

For $i = 0$, clearly  $U_0 = I$. If $i \geq 1$, by induction we know one of (a) or (b) is true for $i-1$. If (a) is true, then we are done. So there is a $U_{i-1} \subseteq I$ with conditions as mentioned earlier. If there was a vertex $v \in V\setminus Q_{i-1}$ with $> p.(hk)^{p-i}$ neighbors in $U_{i-1}$, let $U_i$ to be $U_{i-1} \cap N_G(v)$ and (b) will be true for $i$. Otherwise, all vertices in $V\setminus Q_{i-1}$ have $\leq p.(hk)^{p-i}$ neighbors in $U_{i-1}$, and we do the following:
\begin{itemize}
    \item [] As long as there is an unmarked vertex in $U_{i-1}$, we pick an unmarked vertex $u \in U_{i-1}$ and mark all vertices in $U_{i-1}$ that have a neighbor in $N_G(u)\setminus Q_{i-1}$.  
    %with $v$ out of $Q_{i-1}$.
\end{itemize}
Since $N_G(u) \leq h$ and each vertex in $V \setminus Q_{i-1}$ has $\leq p.(hk)^{p-i}$ neighbors in $U_{i-1}$, at most $ph^{p - i + 1}k^{p-i}$ vertices would get marked after picking $u$. Therefore, we would pick at least $\frac{|U_{i-1}|}{ph^{p - i + 1}k^{p-i}} > \frac{p.(hk)^{p - i + 1}}{ph^{p - i + 1}k^{p-i}} = k$ vertices. Since these vertices are independent, the number of them is at least $k + 1$, they are neighbors of $Q_{i-1}$, and they do not have common neighbors out of $Q_{i-1}$, every subset of them of size $k+1$ forms a \textit{nice} subset and (a) will be true.

If $i = p$, the above proposition implies that either there is a \textit{nice} subset $U\subseteq I$ or a $K_{p, p}$. In the same way as the induction, we also can construct $U_i$ and $Q_i$ using $U_{i-1}$ and $Q_{i-1}$. This is easily doable by checking all  vertices in $V \setminus Q_{i-1}$ to see whether they have $> p.(hk)^{p-i}$ neighbors in $U_{i-1}$. If we could not find such a vertex, then we can find a \textit{nice} subset like the induction by marking vertices. If we could construct all $U_i$s, then we can easily find a $K_{p, p}$ in the induced subgraph of $(U_p \cup Q_p)$.
\end{proof}

\begin{lemma}\label{lem:5}
Let $G = (V, E)$ be a $d$-degenerate graph with possible self-loops. Then there are $\geq \frac{n}{2d + 1}$ vertices $v$ with $|N_G(v)| \leq 2d$.
\end{lemma}
\begin{proof}
Note $\sum_{v \in V} |N_G(v)| = 2(m - \sum_{v \in V} L_G(v)) \leq 2nd$ that the inequality is by Proposition \ref{prop:3}. Suppose number of vertices like $v$ with $|N_G(v)| > 2d$ is $t$. Then we have:
\begin{eqnarray*}
t(2d + 1) &\leq& \sum_{v \in V} |N_G(v)|\ \ \leq\ \ 2nd
\end{eqnarray*}
This implies that 
\begin{eqnarray*}
t &\leq& \frac{2nd}{2d + 1}, \;\; \mbox{and}\\
n - t &\geq& \frac{n}{2d + 1}
\end{eqnarray*}
This completes the proof of the lemma.
\end{proof}

\begin{lemma}\label{lem:6}
Any $d$-degenerate graph $G = (V, E)$ with possible self-loops has an independent set $I$ with size $\geq \frac{n}{(d+1)(2d+1)}$ such that $|N_G(v)| \leq 2d$ for all $v \in I$ and one can find such an independent set in time $(n + m)^{O(1)}$.
\end{lemma}
\begin{proof}
First, construct a $(d+1)$-coloring for $V$ in $n^{O(1)}$ using Proposition \ref{prop:4}. By Lemma \ref{lem:5} there are $\geq \frac{n}{(2d+1)}$ vertices with $|N_G| \leq 2d$ and therefore, there are $\geq \frac{n}{(2d+1)(d+1)}$ vertices with $|N_G| \leq 2d$ and the same color, which means they form an independent set.
\end{proof}

Now, we are ready to describe the kernel. As long as, $n > p(d+1)(2d+1)(2dk)^{p}$, we apply the following reduction rule.

\begin{redrule}
Use Lemma \ref{lem:6} to find an independent set $I$ of size $\geq \frac{n}{(d+1)(2d+1)} > p\cdot(2dk)^p$ such that $|N_G(v)| \leq 2d$ for all $v \in I$. Then, since the given graph $G$ does not contain any $K_{p,p}$, by setting $h = 2d$ and using Lemma \ref{lem:4}, find a \textit{nice} subset $U \subseteq I$. Then remove $u \in U$ that minimizes $|N_G(u)| + L_G(u)$ and add a self-loop to each vertex of $N_G(u)$. 
\end{redrule}

To show the soundness of the reduction rule, we prove the following lemma.

\begin{lemma}\label{lem:7}
Suppose $G=(V, E)$ is a graph with possible self-loops, and $U \subseteq V$ is $nice$. Then, for any $u \in U$ with the minimum $|N_G(u)| + L_G(u)$, there is a solution for {\sc PVC} which does not contain $u$.
\end{lemma}
\begin{proof}
Consider any solution $S$ containing $u$. Since $|S| = k$, there is a $u' \in U$ such that $(i)$ $(\{u'\}\cup PV_U(u')) \cap S=\emptyset$. Therefore, we have:
\begin{align*}
    E_G(S\setminus \{u\} \cup \{u'\}) 
    &\geq E_G(S) - (|N_G(u)| + L_G(u)) + (|N_G(u')| + L_G(u')) \tag{by $(i)$}\\
    &\geq E_G(S) \tag{since $|N_G(u)| + L_G(u) \leq |N_G(u')| + L_G(u')$}
\end{align*}
This implies that $S\setminus \{u\} \cup \{u'\}$ is a solution that does not contain $u$.
\end{proof}

We finally prove Theorem \ref{thrm:2}. For convenience,  we restate the theorem here. 

\thmker*

\begin{proof}
The running time of the described algorithm is polynomial by Lemma \ref{lem:4} and \ref{lem:6}, and the reduction rule is safe by Lemma \ref{lem:7}.
The number of vertices in the kernel is $\leq p(d+1)(2d+1)(2dk)^{p}$, which is $O(pd^2(2dk)^{p})$. Although the number of self-loops may be large, notice that the number of self-loops on a vertex will be at most $n$. We may remove the self-loops and add it as a weight on the vertex. Thus, each weight can be represented using at most $\log n$ bits. Since we have an algorithm for the problem with running time $2^{O(kd)}n^{O(1)}$, i.e., Theorem \ref{thrm:1}, when $kd\leq \log n$ the algorithm runs in polynomial time and thereby, it would be a compression itself. Otherwise, we have that $\log n \leq kd$ that guarantees the weight function $\rho$ mentioned in the theorem statement. 
\end{proof}

\section{Conclusion}
In this work we gave a single exponential parameterized algorithm and a polynomial compression for {\sc PVC} on graphs of bounded degeneracy that include many sparse graph classes like planar graphs and $H$-minor free graphs. 
Is it possible to get similar results on biclique free graphs, a superclass of bounded degeneracy graphs?  
Notice that there is a linear kernel for {\sc Dominating Set} on planar graphs, $H$-minor free graphs, and apex-minor free graphs~\cite{FomilLinKernel,metkernelJACM}. Can we get a linear kernel or compression for {\sc PVC} on planar graphs? 

\bibliographystyle{siam}
\bibliography{ref}
\end{document}